\documentclass[aps,prl,onecolumn,,superscriptaddress,groupedaddress]{revtex4}  
\usepackage{graphicx}  
\usepackage{dcolumn}   
\usepackage{bm}        
\usepackage{amssymb}   
\usepackage{amsmath}
\usepackage{lipsum} 
\usepackage{braket}
\usepackage{cleveref}
\usepackage{siunitx}
\usepackage{xcolor}
\usepackage{mathtools}

\newcommand{\muB}{\mu_{\text{B}}}
\newcommand{\gammaH}{\gamma_{\text{H}}}

\newcommand{\Thf}{T_{\text{hf}}}
\newcommand{\gs}{g_{\text{s}}}
\newcommand{\red}[1]{{\color{red}#1}}
\newcommand*\diff{\mathop{}\!\mathrm{d}}

\begin{document}

\title{Magnetometry with spin polarized Hydrogen from molecular photo-dissociation }

\author{Konstantinos Tazes}
\affiliation{University of Crete, Department of Physics, Heraklion, Greece}
\author{Alexandros K. Spiliotis}
\author{Michalis Xygkis}
\affiliation{Foundation for Research and Technology Hellas, Institute of Electronic Structure and Laser, N. Plastira 100, Heraklion, Crete, Greece, GR-71110}
\affiliation{University of Crete, Department of Physics, Heraklion, Greece}
\author{George E. Katsoprinakis}
\affiliation{Foundation for Research and Technology Hellas, Institute of Electronic Structure and Laser, N. Plastira 100, Heraklion, Crete, Greece, GR-71110}
\author{T. Peter Rakitzis}
\affiliation{Foundation for Research and Technology Hellas, Institute of Electronic Structure and Laser, N. Plastira 100, Heraklion, Crete, Greece, GR-71110}
\affiliation{University of Crete, Department of Physics, Heraklion, Greece}
\author{Georgios Vasilakis}
\affiliation{Foundation for Research and Technology Hellas, Institute of Electronic Structure and Laser, N. Plastira 100, Heraklion, Crete, Greece, GR-71110}

\begin{abstract}

In a recent publication [arXiv:2010.14579], we introduced a new type of atomic magnetometer, which relies on hydrohalide photo-dissociation to create high-density spin-polarized hydrogen. Here, we extend our previous work and present a detailed theoretical analysis of the magnetometer signal and its dependence on time. We also derive the sensitivity for a spin-projection noise limited magnetometer, which can be applied to an arbitrary magnetic field waveform.
\end{abstract}


\maketitle

A broad range of physical objects and processes generate magnetic fields which upon detection can convey important information about the nature and structure of their origin. As a result, magnetic field detection lies at the heart of many scientific and technological applications, which can benefit significantly from advances in magnetometry \cite{book:MagneticSensors}.

Different magnetic field sensors have been developed which offer distinct advantages and are attractive for particular applications.
In general terms, an ideal magnetometer should present high-sensitivity, wide bandwidth detection, high-performance over a large dynamic range and operating conditions, as well as capability for miniaturization when used for magnetic field imaging.

Recently, a new type of atomic magnetometer was demonstrated based on high-density spin-polarized atomic H (SPH) \cite{NanosecondresolvedMagnetometer}, which has the potential to address satisfactorily the above requirements for magnetometry. The spin-polarized ensemble is produced by photo-dissociating hydrohalide gas with a circularly-polarized laser pulse \cite{Rakitzis_Science, Sofikitis_HClHBr_SPH, Rakitzis_ChemPhysChem}. Magnetic field detection is achieved by monitoring the dynamics of the H hyperfine coherences, which are created in the optical pumping process without the need for external magnetic fields. 

This paper is an extension of the work presented in \cite{NanosecondresolvedMagnetometer}, analytically deriving equations for the spin-dynamics, the magnetometer signal and the quantum spin-projection noise.

\begin{figure*}[h]
\includegraphics[width=\textwidth]{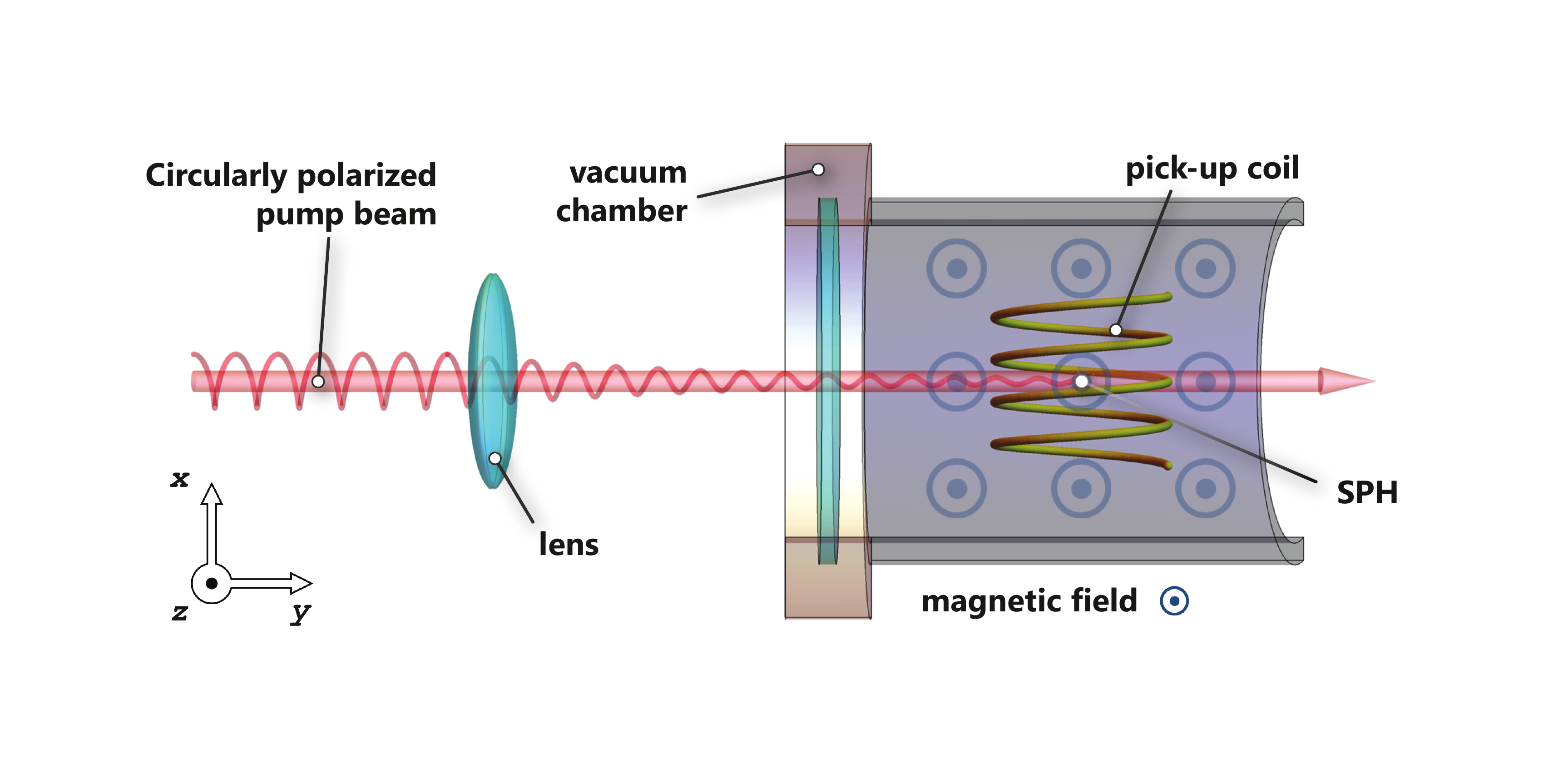}
\caption{Schematic of the magnetometer setup. A circularly-polarized laser pulse (red arrow), photo-dissociate molecular hydrohalide gas producing spin-polarized atomic H. A pickup coil detects the evolution of the ensemble magnetization in the presence of a magnetic field (green arrow).} \label{fig:ExperimentalSetup}
\end{figure*}

We will consider the magnetometer scheme with mutually orthogonal directions for optical pumping, magnetic field direction and spin-probing, as shown in Fig.~\ref{fig:ExperimentalSetup}. Without loss of generality we take the magnetic field to be in the $z$ direction, the optical pumping along the $y$ axis and the probe axis in the $x$ direction. Monitoring of spins is realized with an inductive pick-up coil, which detects the magnetic flux generated by the H spins. Since the electron magnetic moment is more than three orders of magnitude larger than the proton magnetic moment, the coil is to a very good approximation only sensitive to the H electron spins. In the following, we will assume a pickup coil with a response time much shorter than the hyperfine interaction period and neglect complications arising from a non-spherical polarized region or from geometrical factors in the coupling of the magnetic field from spins to the coil. For simplicity, we will assume that the observable is $\frac{d \hat{S}_x}{dt}$, where $\hat{S}_i$ expresses the dimensionless electron spin operator in the $i$ direction.

\section{Time evolution}
In the presence of a magnetic field $\mathbf{B}$ (which can be time varying), the SPH evolves according to the Hamiltonian (we neglect for the moment the relaxation):
\begin{equation}
\hat{H}  = \omega_0 \hat{\mathbf{S}} \cdot \hat{\mathbf{I}} + \frac{\gs \muB}{\hbar} \mathbf{B}  \cdot \hat{\mathbf{S}}= \hat{H}_0+ \frac{\gs \muB}{\hbar} B  \cdot \hat{S}_z \label{eq:S:Hamiltonian0}
\end{equation} 
where $\hat{\mathbf{I}}$ is the dimensionless nuclear spin operator, $\omega_0$ is the hyperfine frequency of H, $\muB$ is the Bohr magneton, $\gs \approx 2$ is the electron spin $g$-factor,  $\hat{H}_0$ is the Hamiltonian for hyperfine interaction expressed in (angular) frequency units. In the above equation, we neglected the coupling of the magnetic field to nuclear spin, as this is three orders of magnitude smaller than the coupling to the electron spin.

For clarity we write the Hamiltonian in the coupled (subscript c) and uncoupled (subscript u) basis taking $z$ (the magnetic field direction) as the quantization axis:
\begin{equation}
\hat{H}_\text{u}=
\begin{pmatrix}
\frac{\omega_0}{4}+\gammaH B & 0 & 0 & 0\\
0 & -\frac{\omega_0}{4}-\gammaH B & \frac{\omega_0}{2} & 0 \\
0 & \frac{\omega_0}{2} & -\frac{\omega_0}{4}+\gammaH B &  0 \\
0 & 0 & 0 & \frac{\omega_0}{4}- \gammaH B
\end{pmatrix}
, \text{ }
\hat{H}_\text{c}=
\begin{pmatrix}
\frac{\omega_0}{4}+\gammaH B & 0 & 0 & 0\\
0 & \frac{\omega_0}{4} & 0 & \gammaH B  \\
0 & 0 & \frac{\omega_0}{4}- \gammaH B &  0 \\
0 & \gammaH B  & 0 & -\frac{3\omega_0}{4}
\end{pmatrix}, \label{eq:S:HamiltonianMatrices0}
\end{equation}
where $\gammaH = \gs \muB / 2\hbar$ is the gyromagnetic ratio of atomic H. The above matrices are expressed in a basis with the following ordering:
\begin{align}
\begin{split}
\text{uncoupled basis: } & \{ |m_s=1/2,m_I =1/2 \rangle , |m_s=-1/2,m_I =1/2 \rangle, |m_s=1/2,m_I = - 1/2 \rangle,|m_s=-1/2,m_I = -1/2 \rangle \},  \\
\text{coupled basis: \phantom{un}} & \{|F=1,m_F =1 \rangle, |F=1,m_F =0 \rangle, |F=1,m_F =-1 \rangle , |F=0,m_F =0 \rangle \}, 
\end{split} \label{eq:S:Basis_Ordering}
\end{align}
where $F$ is the total spin (sum of electronic and nuclear spin) quantum number, $m_s$, $m_I$ and $m_F$ are respectively the electronic, nuclear and total spin projection along the quantization axis.

Transformation of an arbitrary operator $\hat{\mathcal{O}}$ or state vector $|\psi \rangle$ from one basis to the other can be performed according to the following rules:
\begin{equation}
\hat{\mathcal{O}}_\text{c} = \mathcal{T}_{\text{cu}} \cdot \hat{\mathcal{O}}_\text{u} \cdot \mathcal{T}_{\text{cu}}^{-1}, \text{ } \hat{\mathcal{O}}_\text{u} = \mathcal{T}_{\text{cu}}^{-1} \cdot \hat{\mathcal{O}}_\text{c} \cdot \mathcal{T}_{\text{cu}}, \text{ } |\psi \rangle_c = \mathcal{T}_{\text{cu}} |\psi \rangle_\text{u}, \text{ } |\psi \rangle_u = \mathcal{T}_{\text{cu}}^{-1} |\psi \rangle_\text{c}, 
\end{equation} 
\begin{equation}
\mathcal{T}_{\text{cu}} = 
\begin{pmatrix}
1 & 0 & 0 & 0\\
0 & \frac{1}{\sqrt{2}} & \frac{1}{\sqrt{2}} &  0 \\
0 & 0 & 0 &  1 \\
0 & -\frac{1}{\sqrt{2}} & \frac{1}{\sqrt{2}} & 0
\end{pmatrix}.
\end{equation}

The hydrohalide photo-dissociation occurs with a sub-nanosecond laser pulse and optical pumping effectively transfers angular momentum from light to the electronic spin, leaving the nuclear spin degrees of freedom in their thermal (completely unpolarized) state. 
Following the above basis ordering, after optical pumping half of the polarized H atoms are in the quantum state (expressed as a column vector):
\begin{equation}
|\psi(0) \rangle_0 = |\psi_0 \rangle = 
\begin{cases} 
      \left( 1, 0, 0 ,0 \right)_\text{u,y}^{\intercal} \xrightarrow{\hat{\mathcal{R}}_x \cdot \left( 1, 0, 0 ,0 \right)^{\intercal}}   \left( \frac{1}{2}, -\frac{\imath}{2}, -\frac{\imath}{2} ,-\frac{1}{2} \right)_\text{u}^{\intercal} \xrightarrow{\mathcal{T}_{\text{cu}}}   \left( \frac{1}{2}, -\frac{\imath}{2},-\frac{1}{2},0 \right)_\text{c}^{\intercal}  & , \sigma=1 \\
      \left( 0, 0, 0 ,1 \right)_\text{u,y}^{\intercal} \xrightarrow{ \hat{\mathcal{R}}_x \cdot \left( 0, 0, 0 ,1 \right)^{\intercal}}   \left( -\frac{1}{2}, -\frac{\imath}{2}, -\frac{\imath}{2} ,\frac{1}{2} \right)_\text{u}^{\intercal}\xrightarrow{\mathcal{T}_{\text{cu}}}  \left( -\frac{1}{2}, -\frac{\imath}{2}, \frac{1}{2},0 \right)_\text{c}^{\intercal} & , \sigma=-1
   \end{cases},
\end{equation}  
where the subscript y denotes that the quantization axis for spin projections was taken in the $y$ direction (if no axis subscript appears it is implicitly assumed that the quantization axis is in the $z$ direction), $\intercal$ is the transpose operation, $\sigma$ is the helicity of the pumping light pulse and $\hat{\mathcal{R}}_x$ is the rotation matrix around the $x$-axis applied to the uncoupled basis (two spins, each of spin 1/2) : 
\begin{equation}
\hat{\mathcal{R}}_x= e^{-\imath \hat{S}_x \frac{\pi}{2}} \otimes e^{-\imath \hat{S}_x \frac{\pi}{2}}
.
\end{equation}

The other half of the polarized H atoms are in the state:
\begin{equation}
|\psi(0) \rangle_1 = |\psi_1 \rangle = 
\begin{cases} 
      \left( 0, 0, 1 ,0 \right)_\text{u,y}^{\intercal} \xrightarrow{\hat{\mathcal{R}}_x \cdot \left( 0, 0, 1 ,0 \right)^{\intercal}}   \left( -\frac{\imath}{2}, -\frac{1}{2}, \frac{1}{2} ,-\frac{\imath}{2} \right)_\text{u}^{\intercal} \xrightarrow{\mathcal{T}_{\text{cu}}}   \left( -\frac{\imath}{2}, 0,-\frac{\imath}{2},\frac{1}{\sqrt{2}}\right)_\text{c}^{\intercal}  & , \sigma=1 \\
      \left( 0, 1, 0 ,0 \right)_\text{u,y}^{\intercal} \xrightarrow{ \hat{\mathcal{R}}_x \cdot \left( 0, 1, 0 ,0 \right)^{\intercal}}   \left( -\frac{\imath}{2}, \frac{1}{2}, -\frac{1}{2} ,-\frac{\imath}{2} \right)_\text{u}^{\intercal}\xrightarrow{\mathcal{T}_{\text{cu}}}  \left( -\frac{\imath}{2}, 0,-\frac{\imath}{2},- \frac{1}{\sqrt{2}}\right)_\text{c}^{\intercal} & , \sigma=-1
   \end{cases}.
\end{equation} 

For the observable $\frac{d \hat{S}_x}{dt}$ the contribution to signal of atoms initially at state $|\psi(0) \rangle_0$ is a factor of $\gammaH B/\omega_0$ smaller than the contribution of atoms initially at $|\psi(0) \rangle_1$. In the case of $\gammaH B/\omega_0 \ll 1$, the magnetometer signal is mainly determined by the atoms initially at $|\psi(0) \rangle_1$. 

For static magnetic field, Schr\"{o}dinger equation can be solved in a straightforward manner:
\begin{equation}
\imath \frac{\partial}{\partial t}|\psi(t) \rangle = \hat{H} |\psi (t) \rangle \Rightarrow |\psi (t) \rangle = e^{- \imath \hat{H} t } |\psi(0) \rangle 
\end{equation}
where $|\psi(t) \rangle$ is the wavefunction at time $t$ after optical pumping. For $|\psi(0)\rangle=|\psi(0)\rangle_1$ the wavefunction at time $t$ in the coupled basis is:
\begin{align}
|\psi(t) \rangle_1 & = \left( -\frac{1}{2} i e^{-\imath \left(\frac{\omega_0}{4}+\gammaH B \right)  t}, - \imath \sigma \frac{\sqrt{2} \gammaH B e^{\imath \frac{\omega_0}{4} t}\sin \left[ \frac{1}{2} \tilde{\omega} t \right]}{\tilde{\omega}}, -\frac{1}{2} \imath e^{-\imath \left( \frac{\omega_0}{4} - \gammaH B \right)t},
 \sigma \frac{e^{\imath \frac{\omega_0}{4}  t} \left( \imath \omega_0 \sin \left[ \frac{1}{2}\tilde{\omega} t \right] +\tilde{\omega} \cos \left[ \frac{1}{2}\tilde{\omega} t \right]  \right)}{\sqrt{2} \tilde{\omega}} \right)^{\intercal} \\
& \approx -\frac{1}{2}\imath e^{\imath \frac{1}{4}\left (\omega_0+2 \tilde{\omega} \right)t} \left ( e^{-\imath  \left(\frac{\omega_0}{2} + \gammaH B + \frac{1}{2}\tilde{\omega} \right) t},0,
e^{-\imath  \left(\frac{\omega_0}{2} - \gammaH B +\frac{1}{2}\tilde{\omega} \right)t},\imath \sqrt{2} \sigma \right )^{\intercal},
\end{align}
where: $ \tilde{\omega} = \sqrt{\omega_0^2+4 \gammaH^2 B^2}$.

In the general case of a time varying magnetic field the Schr\"{o}dinger equation cannot be solved analytically, as the Hamiltonian does not commute with itself at different times. For an approximate analytical solution it is convenient to work in the interaction picture (denoted by the $\tilde{\phantom{a}}$ symbol):
\begin{equation}
\frac{\partial}{\partial t} |\tilde{\psi}(t) \rangle = -\imath \hat{\tilde{V}}   |\tilde{\psi}(t) \rangle, \label{eq:S:Interaction_Schrodinger}
\end{equation}
where $\hat{\tilde{V}}$ is the Hamiltonian describing the magnetic field coupling to the atoms:
\begin{equation}
\hat{\tilde{V}}(t) = e^{\imath \hat{H}_0 t} \left[ g_s \muB B \hat{S}_z\right] e^{-\imath \hat{H}_0 t} = \gammaH B(t)
\begin{pmatrix}
1 & 0 & 0 & 0\\
0 & - \cos (\omega_0 t) & \imath \sin (\omega_0 t) & 0 \\
0 & -\imath  \sin (\omega_0 t) &  \cos (\omega_0 t) &  0 \\
0 & 0 & 0 & -1
\end{pmatrix}.
\end{equation}
In the last equation $\hat{\tilde{V}}$ is expressed in the uncoupled basis.

The solution of Eq.~\ref{eq:S:Interaction_Schrodinger} can be expressed in the form of the Magnus series \cite{MagnusExpansion}. For magnetic fields that change slowly with respect to the hyperfine frequency so that:
\begin{equation}
\int_0^t B(t') e^{\imath \omega_0 t'} \diff t' \ll \int_0^t B(t') \diff t' \label{eq:S:MagnusConditionApprox}
\end{equation} 
the solution of Eq.~\ref{eq:S:Interaction_Schrodinger} can be approximated by keeping only the first term in Magnus expansion. In this case:
\begin{equation}
|\tilde{\psi}(t) \rangle  \approx \exp \left(-\imath \int_0^t \hat{\tilde{V}}(t') \diff t' \right) |\psi (0) \rangle,
\end{equation}
and the observable (taking into account the condition \ref{eq:S:MagnusConditionApprox}) can be written as:
\begin{align}
\frac{d}{dt} \langle \tilde{\psi}(t) | e^{\imath \hat{H}_0 t}  \hat{S}_x e^{-\imath \hat{H}_0 t} |\tilde{\psi}(t)\rangle & \approx 
\sigma \frac{d}{dt} \left\{ \frac{1}{2} \sin\left[ \gammaH \int_0^t B(t') \diff t' \right] \cos (\omega_0 t) \right\} \\
& \approx -\sigma \frac{\omega_0}{2} \sin\left[ \gammaH \int_0^t B(t') \diff t' \right] \sin (\omega_0 t).
\end{align}
The last approximation holds for $\gammaH B \ll \omega_0$.

The condition stated in \ref{eq:S:MagnusConditionApprox} implies that the magnetic field does not induce hyperfine transitions and the magnetic field can be treated as a perturbation to the energies of the hyperfine levels.  
Then, for small magnetic fields ($\gammaH B \ll \omega_0$) the Hamiltonian for H atoms (hyperfine levels with $F=0$ and $F=1$) can be approximated to be (as before the magnetic field direction is taken to be the quantization axis):
\begin{equation}
\hat{H} \approx \hat{H}_0+\gammaH B(t) (\hat{S}_z+\hat{I}_z). \label{eq:S:ApproxHamiltonian}
\end{equation}
This Hamiltonian commutes with itself at different times and the Schr\"{o}dinger equation can be solved analytically:
\begin{equation}
|\psi (t) \rangle = e^{- \imath \int_0^t \hat{H}(t') \diff t' } |\psi(0) \rangle 
\end{equation} 

\subsection{Decay}
The evolution of spins is also affected by relaxation processes leading to non-Hamiltonian dynamics. We model these by introducing a decay term in the density matrix equation:
\begin{equation}
\frac{d \rho}{dt} = \imath [\rho,H] -  \frac{1}{T_2}(\rho-\rho_{\text{eq}}), \label{eq:S:rhoEvol}
\end{equation}
where $1/T_2$ is the decay rate, and $\rho_{\text{eq}}$ corresponds to the state towards which the decay processes drive the system. We take this quantum state to be the completely unpolarized state, written in the form:
\begin{equation}
\rho_{\text{eq}} = \frac{1}{4} \mathbb{I}_{4 \times 4}, \label{eq:S:rhoEqu}
\end{equation}
where $\mathbb{I}_{4 \times 4}$ is the $4\times4$ identity matrix. For Eqs.~\ref{eq:S:rhoEvol}-\ref{eq:S:rhoEqu} to be valid the population decay (hydrogen atom losses due to formation of molecules) should be significantly slower compared to (atomic) hydrogen spin-decoherence processes. This condition is justified in our case.

The density matrix at $t=0$ is ($z$ quantization axis the direction of magnetic field and optical pumping in the $y$ axis):
\begin{equation}
\rho(0)  = \frac{1}{2}|\psi_0 \rangle \langle \psi_0 | + \frac{1}{2} |\psi_1 \rangle \langle \psi_1 |  \rightarrow
\begin{pmatrix}
\frac{1}{4} & \sigma \frac{\imath}{4} & 0 & 0\\
-\sigma \frac{\imath}{4} & \frac{1}{4} & 0 &  0 \\
0 & 0 & \frac{1}{4} & \sigma \frac{\imath}{4} \\
0 & 0 & -\sigma \frac{\imath}{4} & \frac{1}{4}
\end{pmatrix}_{\text{u}}
\rightarrow \frac{1}{4}
\begin{pmatrix}
1 & \sigma \frac{\imath}{\sqrt{2}} & 0 & - \sigma \frac{\imath}{\sqrt{2}}\\
-\sigma\frac{\imath}{\sqrt{2}} & 1 & \sigma \frac{\imath}{\sqrt{2}} &  0 \\
0 & -\sigma \frac{\imath}{\sqrt{2}} & 1 &  -\sigma\frac{\imath}{\sqrt{2}} \\
\sigma \frac{\imath}{\sqrt{2}} & 0 & \sigma \frac{\imath}{\sqrt{2}} & 1
\end{pmatrix}_{\text{c}}.
\end{equation}

An analytical solution to the density matrix equation can be found for a static magnetic field or for an arbitrary field when the Hamiltonian can be approximated by Eq.~\ref{eq:S:ApproxHamiltonian}. In the experimentally relevant limit of $\omega_0 \gg (1/T2, \gammaH B)$ the signal is (keeping lowest order terms in the harmonic amplitudes):
\begin{equation}
\langle \frac{d \hat{S}_x}{dt} \rangle(t)  = \text{Tr} \left[ \frac{d \rho}{dt} \hat{S}_x \right] \approx -\sigma \frac{1}{4} e^{-t/T_2} \sin \left[ \gammaH \int_0^t B(t') \diff t' \right] \sin (\omega_0 t). \label{eq:S:SignalWithDecay}
\end{equation}

Eqs.~\ref{eq:S:rhoEvol}-\ref{eq:S:rhoEqu} implies spin-damping occurs (at equal rate) for both electron and nuclear spin of H. However, the results derived here are general for the relevant approximations ($\omega_0 \gg (1/T2, \gammaH B$) and condition \ref{eq:S:MagnusConditionApprox}. For instance, Eq.~\ref{eq:S:SignalWithDecay} is reproduced also for decay mechanisms that relax only the electronic spin:
\begin{equation}
\frac{d \rho}{dt} = \imath[\rho,H] -  \frac{4}{3 T_2}(\hat{\mathbf{S}}^2  \cdot \rho-\hat{\mathbf{S}} \cdot \rho \hat{\mathbf{S}} \cdot). \label{eq:S:rhoEvol2}
\end{equation} 

\section{Spin-projection noise}


In the following we will need to know the multi-time correlation
\begin{equation}
\langle \frac{d\hat{S}_x}{dt}(t) \frac{d\hat{S}_x}{dt}(t') \rangle,
\end{equation}
where we take:
\begin{equation}
\frac{d}{dt} \hat{S}_x = \imath \left[ \hat{H}, \hat{S}_x \right]-\frac{1}{T_2} \hat{S}_x 
\end{equation}
The last term (not derived from first principles) was introduced to account for the spin decay.
We assume that the evolution of the density matrix $\rho$ is given by Eq.~\ref{eq:S:rhoEvol} (though the results are the same -within the relevant approximations- for the evolution described in Eq.~\ref{eq:S:rhoEvol2}).

The multi-time correlation can be written operationally in the form:
\begin{equation}
\langle \frac{d\hat{S}_x}{dt}(t) \frac{d\hat{S}_x}{dt}(t') \rangle = \text{Tr} \left \{ \frac{d\hat{S}_x}{dt} \left[  \hat{U}(t,t') \left( \frac{d\hat{S}_x}{dt}  \rho(t') \right) \right] \right \}, \label{eq:S:MultitimeSxDef}
\end{equation}
where $U(t,t')$ is the evolution operator of the density matrix from time $t'$ to $t$ ($t>t'$), and $\rho(t')$ is the density matrix at time $t'$: $\rho(t')=U(t',0) \rho(0)$. The evolution operator $U$ cannot be written in the form of a matrix and is not associative. The evolved state $\hat{U}(t,t') \left( \frac{d\hat{S}_x}{dt}  \rho(t') \right)$ can be found from the general solution of  Eq.~\ref{eq:S:rhoEvol} taking $ \frac{d\hat{S}_x}{dt}  \rho(t')$ as the initial condition for the density matrix.
For quantum noise analysis we can take the magnetic field to be zero in the calculation of the multi-time correlation, since quantum noise affects considerably magnetometry only at low fields.
When $\omega_0 T_2 \gg 1$, it can be found that:
\begin{equation}
\langle \frac{d\hat{S}_x}{dt}(t) \frac{d\hat{S}_x}{dt}(t') \rangle = \frac{1}{8} \omega^2 e^{- |t-t'|/T_2} \cos \left[ \omega_0 (t-t') \right]. \label{eq:S:SignalCor}
\end{equation}

In order to find what is the magnetic field uncertainty due to spin-projection noise, we have to specify a method for estimating the magnetic field from the detected signal $\frac{d \hat{S}_x}{dt}(t)$. Taking into account Eq.~\ref{eq:S:SignalWithDecay}, one way to do this (appropriate for arbitrary magnetic waveforms) is from considering the ``quasi-instantaneous'' amplitude of the frequency component at the hyperfine frequency:
\begin{equation}
\xi(n \Thf) = \frac{1}{\Thf} \int_{n \Thf}^{(n+1) \Thf} \frac{d\hat{S}_x}{dt}(t) \sin (\omega_0 t) \diff t, \label{Eq:S:xi_Def}
\end{equation}
where $\Thf=2 \pi/\omega_0$ is the period of hyperfine oscillation and $n$ is an integer number. We assume that the magnetic field can be written in the form: $B(t)=B_0 \mathcal{K}(t)$, where $\mathcal{K}$ is a known (but other than this an arbitrary), time-dependent function. Spin-projection noise creates an uncertainty in the estimation of $B_0$.

We consider the case where the functions $\mathcal{K}(t)$ and $e^{-t/T_2}$ evolve in time much slower compared to $\sin (\omega_0 t) $ and can therefore be considered constant during the hyperfine period. Effectively this is the situation for $\omega_0 T_2 \gg 1 $ and the condition stated in \ref{eq:S:MagnusConditionApprox}. Then, the quasi-amplitude of the sine wave at hyperfine frequency is (we ignore the -irrelevant for noise purposes- $-\sigma$ factor in the signal in Eq.~\ref{eq:S:SignalWithDecay}):
\begin{equation}
\frac{1}{\Thf} \int_{n \Thf}^{(n+1)\Thf} \frac{d\hat{S}_x}{dt}(t) \sin (\omega_0 t) \diff  t \approx \frac{1}{8}\omega_0  e^{- n \Thf/T_2} \sin \left[ \gamma_H B_0 \int_0^{n \Thf} \mathcal{K} (t') \diff  t' \right] \approx \frac{1}{8}\omega_0 e^{- n \Thf /T_2}  \gamma_H B_0  \int_0^{n \Thf} \mathcal{K}(t') \diff  t',
\end{equation}
where the last approximation holds for small magnetic fields.

The magnetic field $B_0$ can be estimated by minimizing with respect to the parameter $B_0$ the $\chi^2$ function:
\begin{equation}
\chi^2=\sum_{n=0}^M \left[ \frac{1}{8}\omega_0 e^{- n \Thf/T_2} \gamma_H B_0 \int_0^{n \Thf} \mathcal{K}(t') \diff t' - \xi(n \Thf) \right]^2,
\end{equation}
where $M \Thf$ is the total measurement time.
The above equation can be viewed as a curve fitting problem with unknown parameter $B_0$ for a noisy signal.
The solution to the curve fitting problem is:
\begin{equation}
B_0 = \frac{\sum_{n=0}^{M}\xi(n \Thf) e^{- n \Thf / T_2} \int_0^{n \Thf} \mathcal{K}(t') \diff  t' }{\gamma_H \sum_{n=0}^{K} \frac{1}{8}\omega_0 e^{-2 n \Thf/T_2} \left( \int_0^{n \Thf} \mathcal{K}(t') \diff  t' \right)^2},
\end{equation}
and the estimation uncertainty in $B_0$ due to the spin-projection noise is:
\begin{equation}
\delta B_0^2=\frac{\sum_{n'=0}^{M} \sum_{n=0}^{M} \langle \xi(n \Thf) \xi(n' \Thf) \rangle e^{-(n+n') \Thf/T_2} \int_0^{n \Thf} \mathcal{K}(t') \diff t' \int_0^{n' \Thf} \mathcal{K} (t'') \diff  t'' }{\gamma_H^2 \left[ \sum_{n=0}^{K} \frac{1}{8}\omega_0 e^{-2 n \Thf/T_2} \left( \int_0^{n \Thf} \mathcal{K}(t') \diff  t'\right)^2 \right]^2}. \label{eq:S:DeltaB0_0}
\end{equation}
From Eqs.~\ref{eq:S:SignalCor} and \ref{Eq:S:xi_Def} we find (in the limit of $\omega_0 \gg \gamma$):
\begin{equation}
\langle \xi(n \Thf) \xi(n' \Thf) \rangle = \frac{1}{32} \omega_0^2 e^{- \left|  (n-n') \right| \Thf/T_2}, \label{eq:S:xiCor}
\end{equation}
so that:
\begin{equation}
\delta B_0^2 = 2  \frac{\int_0^{T_m} \int_0^{T_m} \diff t \diff t'   e^{- \left (t+ t'\right)/T_2 } e^{- \left | t- t' \right |/T_2 } \int_0^{t} \mathcal{K}(x') \diff  x' \int_0^{t'} \mathcal{K}(x) \diff x }{\left[ \gamma_H  \int_0^{T_m} \diff t  e^{-2 t/T_2} \left( \int_0^{t} \mathcal{K}(x) \diff  x \right)^2 \right]^2}, \label{eq:S:DeltaB0Sq1}
\end{equation}
where $T_m$ is the measurement time for a single run of the experiment.

The above derivation applies to a measurement a single H atom. For $N_{\text{SPH}}$ independent (absence of spin-squeezing) H atoms the uncertainty in magnetic field estimation is:
\begin{equation}
\delta B_0^2 = \frac{2}{N_{\text{SPH}}}  \frac{\int_0^{T_m} \int_0^{T_m} \diff t \diff t'   e^{-  \left (t+ t'\right) /T_2 } e^{- \left | t- t' \right |/T_2 } \int_0^{t} \mathcal{K}(x') \diff  x' \int_0^{t'} \mathcal{K}(x) \diff x }{\left[ \gamma_H  \int_0^{T_m} \diff t  e^{-2 t /T_2} \left( \int_0^{t} \mathcal{K}(x) \diff  x \right)^2 \right]^2}. \label{eq:S:DeltaB0Sq1}
\end{equation}

\section{Summary}
We developed analytical equations describing the operation of the SPH magnetometer shown in Fig.~\ref{fig:ExperimentalSetup}. The time evolution of the quantum state for a time-varying magnetic field was derived and the magnetometer signal was shown to exhibit first order dependence to the sensed field. Finally, the effect of spin-projection noise on the estimation of magnetic field was considered, deriving a formulation for an arbitrary waveform.

\section{Acknowledgements}
This work \red{was} supported by the Hellenic Foundation for Research and Innovation (HFRI) and the General Secretariat for Research and Technology (GSRT), grant agreement No
HFRI-FM17-3709 (project NUPOL) and by the project “HELLAS-CH” (MIS 5002735), which is implemented under the “Action for Strengthening Research and Innovation Infrastructures”, funded by the Operational Programme “Competitiveness, Entrepreneurship and Innovation” (NSRF 2014-2020) and cofinanced by Greece and the European Union (European Regional Development Fund).

\bibliography{Magnetometer_Paper_biblio}

\begin{thebibliography}{1}

\bibitem{book:MagneticSensors}
Pavel Ripka.
\newblock {\em Magnetic Sensors and Magnetometers}.
\newblock Artech House Publishers, 2001.

\bibitem{NanosecondresolvedMagnetometer}
Alexandros~K. Spiliotis, Michail Xygkis, Konstantinos Tazes, George~E.
  Katsoprinakis, Georgios Vasilakis, and T.~Peter Rakitzis.
\newblock A nanosecond-resolved atomic hydrogen magnetometer, 2020.

\bibitem{Rakitzis_Science}
T.~P. Rakitzis, P.~C. Samartzis, R.~L. Toomes, T.~N. Kitsopoulos, Alex Brown,
  G.~G. Balint-Kurti, O.~S. Vasyutinskii, and J.~A. Beswick.
\newblock Spin-polarized hydrogen atoms from molecular photodissociation.
\newblock {\em Science}, 300(5627):1936--1938, 2003.

\bibitem{Sofikitis_HClHBr_SPH}
Dimitris Sofikitis, Luis Rubio-Lago, Lykourgos Bougas, Andrew~J. Alexander, and
  T.~Peter Rakitzis.
\newblock Laser detection of spin-polarized hydrogen from {HCl} and {HBr}
  photodissociation: Comparison of {H}- and halogen-atom polarizations.
\newblock {\em The Journal of Chemical Physics}, 129(14):144302, 2008.

\bibitem{Rakitzis_ChemPhysChem}
Rakitzis~T. P.
\newblock Pulsed-laser production and detection of spin-polarized hydrogen
  atoms.
\newblock {\em Chemphyschem : a European journal of chemical physics and
  physical chemistry}, 5:1489--1494, 2004.

\bibitem{MagnusExpansion}
S.~Blanes, F.~Casas, J.A. Oteo, and J.~Ros.
\newblock The magnus expansion and some of its applications.
\newblock {\em Physics Reports}, 470(5):151 -- 238, 2009.

\end{thebibliography}
\bibliographystyle{unsrt}


\end{document}